# Reactive processing of formaldehyde and acetaldehyde in aqueous aerosol mimics: Surface tension depression and secondary organic products

Z. Li, A. N. Schwier, N. Sareen, and V. F. McNeill*

Department of Chemical Engineering, Columbia University, New York, NY, 10027

*Correspondence to: V. Faye McNeill (vfm2103@columbia.edu)

## Abstract

The reactive uptake of carbonyl-containing volatile organic compounds (cVOCs) by aqueous atmospheric aerosols is a likely source of particulate organic material. The aqueous-phase secondary organic products of some cVOCs are surface-active. Therefore, cVOC uptake can lead to organic film formation at the gas-aerosol interface and changes in aerosol surface tension. We examined the chemical reactions of two abundant cVOCs, formaldehyde and acetaldehyde, in water and aqueous ammonium sulfate (AS) solutions mimicking tropospheric aerosols. Secondary organic products were identified using Aerosol Chemical Ionization Mass Spectrometry (Aerosol-CIMS), and changes in surface tension were monitored using pendant drop tensiometry. Hemiacetal oligomers and aldol condensation products were identified using Aerosol-CIMS. Acetaldehyde depresses surface tension to 65(±2) dyn $cm^{-1}$ in pure water (a 10% surface tension reduction from that of pure water) and 62(±1) dyn $cm^{-1}$ in AS solutions (a 20.6% reduction from that of a 3.1 M AS solution). Surface tension depression by formaldehyde in pure water is negligible; in AS solutions, a 9% reduction in surface tension is observed. Mixtures of these species were also studied in combination with methylglyoxal in order to evaluate the influence of cross-reactions on surface tension depression and product formation in these systems. We find that surface tension depression in the solutions containing mixed cVOCs exceeds that predicted by an additive model based on the single-species isotherms.

# 1 Introduction

Organic material is a ubiquitous component of atmospheric aerosols, making up a major fraction of fine aerosol mass, but its sources and influence on aerosol properties are still poorly constrained (Jimenez et al., 2009; Kanakidou et al., 2005). Many common organic aerosol species are surface-active (Facchini et al., 1999; Shulman et al., 1996). Surface-active molecules in aqueous solution form structures that allow hydrophobic groups to avoid contact with water while hydrophilic groups remain in solution. In an aqueous aerosol particle, they may partition to the gas-aerosol interface, reducing aerosol surface tension and potentially acting as a barrier to gas-aerosol mass transport (Folkers et al., 2003; McNeill et al., 2006). Depressed aerosol surface tension due to film formation may lead to a decrease in the critical supersaturation required for the particle to activate and grow into a cloud droplet as described by Köhler Theory (Kohler, 1936). The surface tension of atmospheric aerosol samples tends to be lower than that predicted based on the combined effects of the individual surfactants identified in the aerosol (Facchini et al., 1999). This is in part because some surface-active aerosol organics remain unidentified. Additionally, the effects of interactions among these species under typical aerosol conditions (i.e. supersaturated salt concentrations, acidic, multiple organic species) are generally unknown.

The adsorption of volatile organic compounds (VOCs) to aqueous aerosol and cloud droplet surfaces has been proposed as a route for the formation of organic surface films (Djikaev and Tabazadeh, 2003; Donaldson and Vaida, 2006). There is also growing evidence that the reactive uptake of the carbonyl-containing VOCs (cVOCs) methylglyoxal and glyoxal by cloud droplets or aerosol water, followed by aqueous-phase chemistry to form low-volatility products, is a source of secondary organic aerosol material (Ervens and Volkamer, 2010; Lim et al., 2010). We recently showed that methylglyoxal suppresses surface tension in aqueous aerosol mimics (Sareen et al., 2010).

Formaldehyde and acetaldehyde, two abundant, highly volatile aldehydes, can be directly emitted from combustion and industrial sources or generated *in situ* via the oxidation of other VOCs (Seinfeld and Pandis, 1998). In aqueous solution, both formaldehyde and acetaldehyde become hydrated and form acetal oligomers, similar to methylglyoxal and glyoxal (Loudon, 2009) (see Figure S2 in the Supporting Information for a schematic of the different reaction mechanisms discussed in this study). Nozière and coworkers showed that acetaldehyde forms light-absorbing aldol condensation products in aqueous ammonium sulfate solutions (Noziere et al., 2010a). Formaldehyde was also recently suggested to react with amines to form organic salts in tropospheric aerosols (Wang et al., 2010). Due to their prevalence and known aqueous-phase oligomerization chemistry, the reactive processing of these species in aqueous aerosol mimics, alone and in combination with other cVOCs, is of interest, but has not been thoroughly studied to date.

We investigated the chemical reactions of formaldehyde and acetaldehyde in pure water and concentrated ammonium sulfate (AS) solutions mimicking aerosol water. The potential of these species to alter aerosol surface tension was examined, and secondary organic products were identified using Aerosol Chemical Ionization Mass Spectrometry (Aerosol-CIMS).

## 2 Experimental Methods

Aqueous solutions containing varying concentrations of organic compounds (acetaldehyde, formaldehyde and/or methylglyoxal) with near-saturation concentrations (3.1 M) of AS were prepared in 100 mL Pyrex vessels using Millipore water. The concentration of formaldehyde used was 0.015 – 0.21 M. The concentration of acetaldehyde was 0.018 M – 0.54 M. In the preparations, 5 mL ampules of 99.9 wt% acetaldehyde (Sigma Aldrich) were diluted to 1.78 M using Millipore water immediately after opening in order to minimize oxidization. Varying amounts of this stock solution were used to prepare the final solutions within 30 minutes of opening the ampule. Formaldehyde and methylglyoxal (MG) were introduced from 37 wt% and 40 wt% aqueous solutions (Sigma Aldrich), respectively. The pH value of the reaction mixtures, measured using a digital pH meter (Accumet, Fisher Scientific), was 2.7-3.1. The acidity of the solutions is attributable to trace amounts of acidic impurities within the organic reagent stock solutions (i.e. pyruvic acid from the MG stock solution).

The surface tension of each sample was measured 24 hours after solution preparation using pendant drop tensiometry (PDT). Pendant drops were suspended from the tip of glass capillary tubes using a 100 μL syringe. The images of the pendant drops were captured and analyzed to determine the shape factor, *H*, and equatorial diameter, $d_e$, as described previously (Sareen et al., 2010; Schwier et al., 2010). These parameters were used to calculate the surface tension according to:

$$\sigma = \frac{\Delta\rho g d_e^2}{H} \quad (1)$$

where $\sigma$ is surface tension, $\Delta\rho$ is the difference in density between the solution and the gas phase, and $g$ is acceleration due to gravity (Adamson and Gast, 1997). Solution density was measured using an analytical balance (Denver Instruments). The drops were allowed to equilibrate for 2 minutes before image capture. Each measurement was repeated 7 times.

Aerosol-CIMS was used to detect the organic composition of the product mixtures as described in detail previously (Sareen et al., 2010; Schwier et al., 2010). Mixtures of formaldehyde, acetaldehyde-MG, and formaldehyde-MG in water and 3.1 M AS were prepared. Total organic concentration ranged from 0.2-2 M. All the solutions containing AS were diluted after 24 hours with Millipore water until the salt concentration was 0.2 M. The solutions were aerosolized in a stream of $N_2$ using a constant output atomizer (TSI) and flowed through a heated 23 cm long, 1.25 cm ID PTFE tube (maintained at 135°C) at RH >50% before entering the CIMS, in order to volatilize the organic species into the gas phase for detection. The time between atomization and detection is ≤ 3.5 s. Since the timescale for the oligomerization of these organics is on the order of hours (Sareen et al., 2010; Nozière et al., 2010a) the detected molecules are most likely formed in the bulk aqueous solutions. The solutions were tested in both positive and negative ion mode, using $H_3O^+\cdot(H_2O)_n$ and $I^-$ as reagent ions, respectively. The applicability of this approach to the detection of acetal oligomers and aldol condensation products formed by dicarbonyls in aqueous aerosol mim-ics has been demonstrated previously (Sareen et al., 2010; Schwier et al., 2010). The average particle concentration was ~$4\times10^4$ $cm^{-3}$ and the volume weighted geometric mean diameter was 414(±14) nm.

The Pyrex vessels shielded the reaction mixtures from UV light with wavelengths < 280 nm (Corning, Inc.), but the samples were not further protected from visible light. We previously showed that exposure to visible light in identical vessels does not impact chemistry in the glyoxal-AS or MG-AS reactive systems (Sareen et al., 2010; Shapiro et al., 2009).

# 3 Results

## 3.1 Surface Tension Measurements

**3.1.1 Single-organic mixtures**. Results of the PDT experiments (Fig. 1) show that both formaldehyde and acetaldehyde depress surface tension in 3.1 M AS solution, but the formaldehyde mixture is less surface-active than that of acetaldehyde. The formaldehyde-AS solutions reach a minimum surface tension of 71.4±0.4 dyn $cm^{-1}$ (a 9% reduction from that of a 3.1 M AS solution ($\sigma$ =78.5±0.3 dyn $cm^{-1}$)), at 0.082 mol C/kg $H_2O$. The acetaldehyde-AS solutions showed more significant surface tension depression. The surface tension of the solutions reached a minimum of 62±1 dyn $cm^{-1}$ (a 20.6% reduction compared to 3.1 M AS solution), when the acetaldehyde concentration exceeded 0.527 mol C/kg $H_2O$. Compared to the surface tension of the acetaldehyde in 3.1 M AS, the surface tension depression of acetaldehyde in water is less significant. The surface tension of acetaldehyde in water decreases rapidly and reaches a minimum value of 65±2 dyn $cm^{-1}$ at 0.89 mol C/kg $H_2O$ (a 10% reduction from that of pure water, 72 dyn $cm^{-1}$). Formaldehyde does not show any detectable surface tension depression in water in the absence of AS.

The surface tension data can be fit using the Szyszkowski-Langmuir equation:

$$\sigma = \sigma_0 - aT\ln(1 + b\mathrm{C}) \quad (2)$$

where $\sigma$ and $\sigma_0$ are surface tension of the solution with and without organics, $T$ is ambient temperature (298 K), C is total organic concentration (moles carbon per kg $H_2O$), and $a$ and $b$ are fit parameters (Adamson and Gast, 1997). The parameters from the fits to the data in Fig. 1 are listed in Table 1.

**3.1.2 Binary mixtures.** Surface tension results for aqueous solutions containing a mixture of two organic compounds (MG and formaldehyde or acetaldehyde) and 3.1 M AS are shown in Fig. 2. For a given total organic concentration (0.5 or 0.05 M), the surface tension decreased with increasing MG concentration. Re-plotting the data from Fig. 2 as a function of MG concentration, it is apparent that the surface tension was very similar for mixtures with the same MG concentration, regardless of the identity or amount of the other species present in the mixture (Fig. 3).

Henning and coworkers developed the following model based on the Szyszkowski-Langmuir equation to predict the surface tension of complex, nonreacting mixtures of organics (Henning et al., 2005):

$$\sigma = \sigma_0(T) - \sum_i \chi_i a_i T \ln(1 + b_\mathrm{i} \mathrm{C}_i) \quad (3)$$

Here, $\mathrm{C}_i$ is the concentration of each organic species (moles carbon per kg $H_2O$), $\chi_i$ is the concentration (moles carbon per kg $H_2O$) of compound $i$ divided by the total soluble carbon concentration in solution, and $a_i$ and $b_i$ are the fit parameters from the Szyszkowski-Langmuir equation for compound $i$. The Henning model has been shown to describe mixtures of nonreactive organics, such as succinic acid-adipic acid in inorganic salt solution, well (Henning et al., 2005). We also found that it was capable of describing surface tension depression in reactive aqueous mixtures containing MG, glyoxal, and AS (Schwier et al., 2010).

The predicted surface tension depression for the binary mixtures as calculated with the Henning model is shown in Fig. 2 as a black line, and the confidence intervals based on uncertainty in the Szyszkowski-Langmuir parameters are shown in grey. The experimentally measured surface tensions are, in general, lower than the Henning model prediction, indicating a synergistic effect between MG and acetaldehyde/formaldehyde. The error of the prediction for the mixtures of MG and acetaldehyde is between 8-24%. The error tends to increase with the concentration of MG. However, the error is less than 10% for formaldehyde-MG mixtures.

**3.1.3 Ternary mixtures.** As shown in Fig. 4, 3.1 M AS solutions containing ternary mixtures of MG, acetaldehyde and formaldehyde also exhibit surface tension depression lower than that predicted by the Henning model. For the ternary mixture experiments, the molar ratio of acetaldehyde to formaldehyde was either 1:3 (Fig. 4a and 4b) or 1:1 (Fig. 4c and 4d) and the MG concentration was varied. The total organic concentration remained constant at 0.05 M. Recasting the data of Fig. 4 as a function of MG concentration shows a similar trend as what was observed for the binary mixtures; for a constant total organic concentration, MG content largely determines the surface tension, regardless of the relative amounts of acetaldehyde and formaldehyde present (Fig. 3).

## 3.2 Aerosol-CIMS characterization

The CIMS data show products of self- and cross-reactions of formaldehyde, acetaldehyde and MG in pure water and 3.1 M AS. The resolution for all CIMS data presented here was m/z ±1.0 amu. All the peaks identified and discussed in the following sections have signal higher than that present in a $N_2$ background spectrum. Any unlabeled peaks are within the background, and were not included in the peak assignment analysis. We did not perform Aerosol-CIMS analysis on acetaldehyde-AS or acetaldehyde-$H_2O$ solutions because these systems have been characterized extensively by others (Casale et al., 2007; Noziere et al., 2010a). These studies showed the acid-catalyzed formation of aldol condensation products in solutions containing AS.

**3.2.1 Formaldehyde.** The mass spectra for formaldehyde in $H_2O$ and in 3.1 M AS obtained using negative ion detection with $I^-$ as the reagent ion is shown in Fig. 5. Possible structures are shown in Table 2. The spectrum shows peaks with mass-to-charge ratios corresponding to formic acid at 81.7 ($CHO_2^-\cdot 2H_2O$) and 208.7 amu ($I^-\cdot CH_2O_2\cdot 2H_2O$) and several peaks consistent with hemiacetal oligomers. 223.3, 291.1, and 323.5 amu are consistent with clusters of hemiacetals with $I^-$. The peaks at 95.6, 110.4, 273.8 and 304.7 amu are consistent with clusters of ionized hemiacetals with $H_2O$. While ionization of alcohols by $I^-$ is normally not favorable, ionized paraformal-dehyde-type hemiacetals are stabilized by interactions between the ionized ─$O^-$ and the other terminal hydroxyl group(s) on the molecule (see the Supporting Information).

The peaks at m/z 176.7 and 193.8 amu are not observed in the formaldehyde-$H_2O$ spectrum, implying that the species observed at those masses are formed via reaction with AS. Within our instrument resolution, these peaks could be consistent with methanol, which is present in our system due to its use as a stabilizer in formaldehyde solutions. However, methanol does not form stable clusters with $I^-$ and therefore will not be detected using this ionization scheme. The peak at 193.8 amu is consistent with an organosulfate species formed from a formaldehyde hemiacetal dimer ($C_2H_5O_6S^- \cdot 2H_2O$) and a satellite peak is also visible at 195.6 amu (see Supporting Information). The abundance of these peaks should be consistent with a 96:4 ratio of stable sulfate isotopes ($^{32}S$ and $^{34}S$), and instead this ratio is found to be 86:14. This is not inconsistent with the identification of an organosulfate species at 193.8 amu, but additional compounds could also be present at 195.6 amu. The peak at 176.7 amu matches an ion formula of $C_6H_9O_6^-$, but the structure and formation mechanism is unknown. Future mechanistic studies are needed in order to resolve products such as this one with unknown chemical structures and/or formation mechanisms.

The positive-ion spectrum of the formaldehyde solution in 3.1 M AS corroborates the identification of hemiacetal oligomers. The formaldehyde hemiacetal dimer sulfate was not observed in positive-ion mode. This was expected, since, to our knowledge, organosulfate species have not previously been observed using positive-ion-mode mass spectrometry (Sareen et al., 2010). The spectrum and peak assignments can be found in the Supporting Information.

**3.2.2 Formaldehyde-methylglyoxal mixtures.** The negative-ion spectrum (detected with $I^-$) for an aqueous mixture of formaldehyde, MG, and AS is shown in Fig. 6, with peak assignments listed in Table 3. Most of the peaks are consistent with formaldehyde hemiacetal oligomers, such as 186.7, 203.5, 230.3, 257.4, and 264.5 amu. Formic acid was detected at 172.8 amu and 208.7 amu. The peak at 288.1 corresponds to MG self-reaction products formed either via aldol condensation or hemiacetal mechanisms (Sareen et al., 2010; Schwier et al., 2010). Several peaks could correspond to self-reaction products of either formaldehyde or MG: 216.5, 252.4, 324.5, and 342.6 amu. The peak at 314.3 amu is consistent with a hemiacetal oligomer formed via cross-reaction of MG with two formaldehyde molecules, clustered with $I^-$ and two water molecules. The peak at 272.2 amu could correspond to either a similar cross-reaction product (MG + 2 formaldehyde) or a MG dimer. Formaldehyde hemiacetal self-reaction products and formic acid were detected in the positive-ion spectrum (Supporting Information).

While the negative-ion spectra of the formaldehyde-AS and formaldehyde-MG-AS mixtures share many similar peaks, there are some differences in the spectra. Small variations in pressure and flow rates within the declustering region can affect the clustering efficiency between the analyte and the parent ions, and surrounding water molecules, resulting in the same analyte compound appearing at different m/z values.

**3.2.3 Acetaldehyde-methylglyoxal mixtures.** The $H_3O^+(H_2O)_n$ spectrum for aqueous acetaldehyde-MG-AS mixtures is shown in Figure 7, with peak assignments listed in Table 4. Several peaks, specifically acetaldehyde aldol condensation products (i.e. 88.9, 107.2, 192.9, 289.6, and 297 amu), are similar to those expected in acetaldehyde-AS solutions (Casale et al., 2007; Noziere et al., 2010a). Hydrated acetaldehyde can be observed at 98.4 amu. Several peaks are consistent with the cross-reaction products of MG and acetaldehyde via an aldol mechanism (126.0, 134.0, 206.7, and 248.9 amu). Formic, glyoxylic, and glycolic acids correspond to the peaks at 84.4, 93.5, and 95.5 amu, respectively. A trace amount of formic acid impurity exists in the 37% formaldehyde aqueous stock solution. Since no significant source of oxidants exists in the reaction mixtures, the formation mechanisms for these species in this system are unknown. The peaks at 88.9 and 107.2 are consistent with either pyruvic acid or crotonaldehyde. Large aldol condensation products from the addition of 6-10 acetaldehydes are observed at 192.9, 289.6, and 297 amu. The peaks at 145.1, 162.9, 164.7 and 235 amu are consistent with MG self-reactions, as discussed by Sareen et al (2010). The peak at 137.3 amu is consistent with a species with molecular formula $C_5H_{12}O_3$, but the mechanism is unknown.

The $I^-$ negative-ion spectrum for acetaldehyde-MG-AS mixtures shows similar results to the positive-ion spectrum (see Figure 8 and Table 5), however aldol condensation products are not detected by this method unless they contain a terminal carboxylic acid group or neighboring hydroxyl groups (Sareen et al., 2010). Small acid species, such as formic, acetic and crotonic acid (172.7 (208.4), 186.4 and 230.7 amu, respectively), were detected. Hydrated acetaldehyde (189.6 and 224.1 amu) and MG (216.3 amu), and hemiacetal self-dimers of acetaldehyde and MG (230.7, 256.4, 264.4, 269.5, and 342.3) were also observed. 256.4 amu is consistent with a MG aldol condensation dimer product, and 272.2 amu could correspond either to a MG hemiacetal dimer or an aldol condensation product. 242.9 amu, $I^-\cdot C_5H_8O_3$, is consistent with an aldol condensation cross product of MG and acetaldehyde. 194.6 amu corresponds to $C_6H_9O_6^-$ ( mechanism and structure unknown).

Note that several peaks appear at similar mass-to-charge ratios in the negative mode mass spectra of both the formaldehyde-MG and acetaldehyde-MG mixtures. MG self-reaction products are expected to be present in both systems. Beyond this, formaldehyde and acetaldehyde are structurally similar small molecules which follow similar oligomerization mechanisms alone and with MG. In several cases, peaks in the mass spectra corresponding to structurally distinct expected reaction products for each system have similar mass-to-charge ratios. For example, the formaldehyde hemiacetal 4-mer ($I^- C_4H_{10}O_5$) and the acetaldehyde dimer ($I^- C_4H_6O_3 \cdot 2H_2O$) are both apparent at 264 amu.

## 4. Discussion

Both formaldehyde and acetaldehyde, and their aqueous-phase reaction products, were found to depress surface tension in AS solutions. However, surface tension depression was not observed in aqueous formaldehyde solutions containing no salt, due to the hydrophilic character of hydrated formaldehyde and its oligomer products. Net surface tension depression by acetaldehyde was greater in the AS solutions than in pure water. These differences for both organics are likely due to chemical and physical effects of "salting out" (Setschenow, 1889), which may enhance organic film formation on the surface of a pendant drop (or aerosol particle). The salt promotes the formation of surface-active species: several of the reaction products in the AS systems identified using Aerosol-CIMS are known or expected to be surface-active, such as organosulfates (Noziere et al., 2010b) and organic acids. Salts can also alter the partitioning of these volatile yet water-soluble organic species between the gas phase and aqueous solution. Formaldehyde has a small Henry's Law constant of 2.5 M $atm^{-1}$, although hydration in the aqueous phase leads to an effective Henry's Law constant of $3\times10^{3}$ M $atm^{-1}$, similar to that of MG (Betterton and Hoffmann, 1988; Seinfeld and Pandis, 1998). The effective Henry's Law constant for acetaldehyde in water at 25°C was measured by Betterton and Hoffmann (1988) to be 11.4 M $atm^{-1}$. The Henry's Law constant of formaldehyde was shown by Zhou and Mopper to increase slightly in aqueous solutions containing an increasing proportion of seawater (up to 100%), but the opposite is true for acetaldehyde (Zhou and Mopper, 1990). The reaction mixtures studied here equilibrated with the headspace of the closed container for 24 h before the surface tension measurements were performed. Each pendant drop equilibrated for 2 min before image capture, after which time there was no detectable change in drop shape. Some of the organics may be lost to the gas phase during equilibration. However, the lower volatility of the aqueous-phase reaction products, especially those formed through oligomerization, leads to significant organic material remaining in the condensed phase (enough to cause surface tension depression and be detected via Aerosol-CIMS).

When formaldehyde and acetaldehyde are present in combination with MG, as would likely happen in the atmosphere (Fung and Wright, 1990; Grosjean, 1982; Munger et al., 1995), there is a synergistic effect: surface tension depression in the solutions containing mixed organics exceeds that predicted by an additive model based on the single-species isotherms. This effect could be due to the formation of more surface-active reaction products in the mixed systems. The deviation from the Henning model prediction was less than 10% except in the case of the acetaldehyde-MG-AS mixtures. Between 21-30% of the detected product mass was identified as cross products in the Aerosol-CIMS positive mode analysis of the acetaldehyde-MG mixtures following (Schwier et al., 2010). Most of the oligomers identified in this system were aldol condensation products, which have fewer hydroxyl groups than acetal oligomers and are therefore expected to be more hydrophobic. A number of organic acid products, likely to be surface-active, were also identified in the acetaldehyde-MG-AS system.

In contrast to the MG-glyoxal system, the presence of formaldehyde and/or acetaldehyde in aqueous MG-AS solutions does influence surface tension depression, in fact, to a greater extent than predicted by the Henning model. However, the results of the binary and ternary mixture experiments suggest that MG still plays a dominant role in these systems since the measured surface tension was remarkably similar in each mixture for a given MG concentration.

The formaldehyde hemiacetal dimer sulfate ($C_2H_6O_6S$) may form via the reaction of $C_2H_6O_3$ with $H_2SO_4$ (Deno and Newman, 1950) (see the Supporting Information for detailed discussion and calculations). The equilibrium concentration of $H_2SO_4$ in our bulk solutions (3.1 M AS, pH = 3) is small ($2.8\times10^{-7}$ M). Minerath and coworkers showed that alcohol sulfate ester formation is slow under tropospheric aerosol conditions (Minerath et al., 2008). Based on our observations, assuming a maximum Aerosol-CIMS sensitivity of 100 Hz ppt$^{-1}$ to this species (Sareen et al., 2010) we infer a concentration of $\geq2\times10^{-4}$ M in the bulk solution after 24 h of reaction. Using our experimental conditions and the kinetics of ethylene glycol sulfate esterification from Minerath et al., we predict a maximum concentration of $7\times10^{-8}$ M. This disagreement between model and experiment suggests that either a) the kinetics of sulfate esterification for paraformaldehyde are significantly faster than for alcohols b) $SO_4^{-2}$ or $HSO_4^-$ is the active reactant, contrary to the conclusions of Deno and Newman, or c) sulfate esterification is enhanced by the solution dehydration that accompanies the atomization and volatilization steps in our detection technique. Photochemical production of organosulfates has also been observed (Galloway et al., 2009; Noziere et al., 2010b; Perri et al., 2010). Our samples were protected from UV light by the Pyrex reaction vessels, and no significant OH source was present, so we don't expect photochemical organosulfate production to be efficient in this system.

Nitrogen-containing compounds could also be formed in these reaction mixtures due to the presence of the ammonium ion (Galloway et al., 2009; Noziere et al., 2009; Sareen et al., 2010). No unambiguous identifications of C-N containing products were made in this study, but analysis using a mass spectrometry technique with higher mass resolution could reveal their presence.

Ambient aerosol concentrations of formaldehyde and acetaldehyde have been measured up to 0.26 μg $m^{-3}$ formaldehyde and 0.4 μg $m^{-3}$ acetaldehyde in Los Angeles (Grosjean, 1982). Using a dry aerosol mass of 50 μg $m^{-3}$, at a relative humidity of 80% (with a mass ratio of water:solute of 1), these ambient in-particle concentrations of formaldehyde and acetaldehyde correspond to 0.17 and 0.18 mol/kg $H_2O$, respectively, which are within the concentration ranges used in this study. At these realistic concentrations, we observed non-negligible surface tension depression by formaldehyde and acetaldehyde (8.8% and 12.1%, respectively). However, if we assume a relative humidity of 99%, relevant for cloud droplet activation, the mass ratio of water:solute increases to 35, so the in-particle concentrations correspond to 0.0049 and 0.0052 mol/kg $H_2O$, respectively, which is below the concentration ranges used. Relatively high aldehyde concentrations are considered justified to mimic the aerosol phase in experiments, which was our intent here (Ervens et al., 2011; Sareen et al., 2010; Tan et al., 2009; Tan et al., 2010). Furthermore, the extended concentration range used here was chosen to enable us to characterize the surface tension behavior using the Szyszkowski-Langmuir equation.

The relatively small Henry's Law partitioning of formaldehyde and acetaldehyde to water suggests that their potential to contribute to total SOA mass is low as com-pared to highly soluble species such as glyoxal. This is supported by the observations of Kroll et al. (2005) that AS aerosols exposed to formaldehyde in an aerosol reaction chamber did not result in significant particle volume growth. However, recent studies have indicated that aldehydes partition into the aqueous particle phase more than predicted by Henry's Law (Baboukas et al., 2000; Grosjean, 1982; Healy et al., 2008); this is hypothesized to be a hydration equilibrium shift (Yu et al., 2011). Grosjean et al. determined that in-particle concentrations were up to 3 orders of magnitude higher for formaldehyde than those predicted by Henry's Law (using an aerosol mass 150 μg $m^{-3}$ and 15% water content). Additionally, formaldehyde and acetaldehyde in the gas phase could adsorb at the aerosol surface (vs. bulk aqueous absorption), and this may also impact aerosol surface tension (Donaldson and Vaida, 2006). Furthermore, Romakkaniemi and coworkers recently showed significant enhancement of aqueous-phase SOA production by surface-active species when OH oxidation is also occurring, beyond what would be predicted based on Henry's Law due to surface-bulk partitioning (Romakkaniemi et al., 2011).

## 5. Conclusions

Two highly volatile organic compounds, formaldehyde and acetaldehyde, were found to form secondary organic products in aqueous ammonium sulfate (AS) solutions mimicking tropospheric aerosols. These species, and their aqueous-phase reaction products, lead to depressed surface tension in the aqueous solutions. This adds to the growing body of evidence that VOCs are a secondary source of surface-active organic material in aerosols.

## Acknowledgement

This work was funded by the NASA Tropospheric Chemistry program (grant NNX09AF26G) and the ACS Petroleum Research Fund (Grant 48788-DN14). The authors gratefully acknowledge the Koberstein group at Columbia University for use of the pendant drop tensiometer.

**Table 1.** Szyszkowski-Langmuir Fit Parameters according to Eq. (2)

| **Mixture** | $\sigma_0$ **(dyn cm$^{-1}$)** | ***a*** **(dyn cm$^{-1}$ K$^{-1}$)** | ***b*** **(kg $H_2O$ (mol C)$^{-1}$)** |
|---|---|---|---|
| Methylglyoxal + 3.1 M $(NH_4)_2SO_4$ (Sareen et al. 2010) | 78.5 | 0.0185±0.0008 | 140±34 |
| Acetaldehyde + 3.1 M $(NH_4)_2SO_4$ | 78.5 | 0.0008±0.0046 | 9.53±3.86 |
| Formaldehyde + 3.1 M $(NH_4)_2SO_4$ | 78.5 | 0.0119±0.0043 | 50.23±44.8 |
| Acetaldehyde + $H_2O$ | 72.0 | 0.0037±0.0011 | 491.64±689 |

**Table 2.** Proposed peak assignments for Aerosol-CIMS mass spectra with $I^-$ of atomized solutions of 0.2 M formaldehyde in 3.1 M AS.

| m/z (amu) ± 1.0 amu | Ion Formula | Molecular Formula | Possible Structures | Mechanism |
|---|---|---|---|---|
| 81.7 | $CHO_2^-\cdot 2H_2O$ | $CH_2O_2$ | 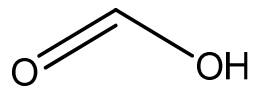 | Formic Acid |
| 95.6 | $C_2H_5O_3^-\cdot H_2O$ | $C_2H_6O_3$ | 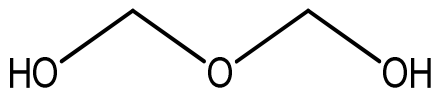 | n=2 hemiacetal |
| 110.4 | $C_2H_3O_3^-\cdot 2H_2O$ | $C_2H_4O_3$ | 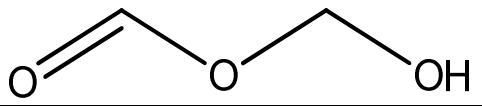 | n=2 hemiacetal |
| 176.7 | $C_6H_9O_6^-$ | $C_6H_{10}O_6$ | Unknown | Unknown |
| 193.8 | $C_2H_5O_6S^-\cdot 2H_2O$ | $C_2H_6O_6S$ | 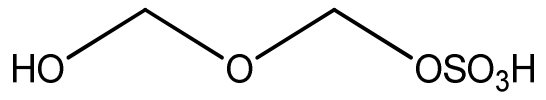 | Hemiacetal sulfate |
| 208.7 | $I^-\cdot CH_2O_2\cdot 2H_2O$ | $CH_2O_2$ | 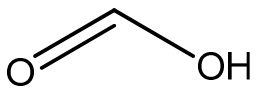 | Formic Acid |
| 223.3 | $I^-\cdot C_2H_6O_3\cdot H_2O$ | $C_2H_6O_3$ | 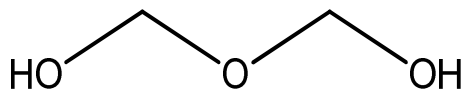 | n=2 hemiacetal |
| 273.8 | $C_8H_{15}O_9^-\cdot H_2O$ <br> $C_8H_{17}O_{10}^-$ | $C_8H_{16}O_9$ <br> $C_8H_{18}O_{10}$ | 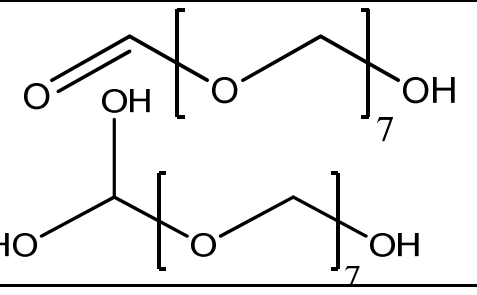 | n=8 hemiacetal |
| 291.1 | $I^-\cdot C_5H_8O_6$ | $C_5H_8O_6$ | 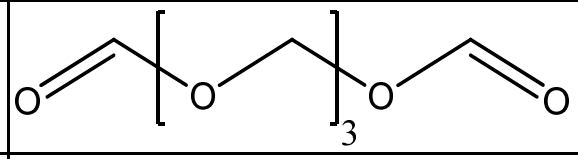 | n=5 hemiacetal |
| 304.7 | $C_9H_{19}O_{10}^-\cdot H_2O$ | $C_9H_{20}O_{10}$ | 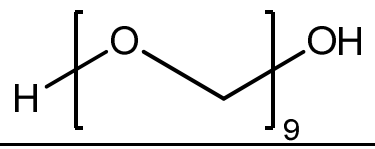 | n=9 hemiacetal |
| 323.5 | $I^-\cdot C_6H_{14}O_7$ | $C_6H_{14}O_7$ | 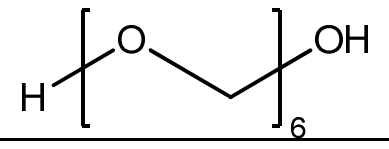 | n=6 hemiacetal |

**Table 3.** Proposed peak assignments for Aerosol-CIMS mass spectra with $I^-$ of atomized solutions of 2 M formaldehyde/MG (1:1) in 3.1 M AS.

| m/z (amu) ± 1.0 amu | Ion Formula | Molecular Formula | Possible Structures | Mechanism |
|---|---|---|---|---|
| 172.8 | $I^-\cdot CH_2O_2$ | $CH_2O_2$ | 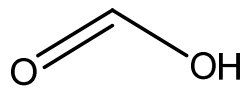 | Formic Acid |
| 186.7 | $I^-\cdot C_2H_4O_2$ | $C_2H_4O_2$ | 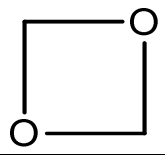 | cyclic F acetal |
| 203.5 | $C_5H_{11}O_6^-\cdot 2H_2O$ | $C_5H_{12}O_6$ | 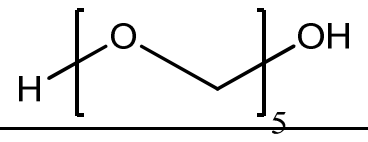 | n=5 F hemiacetal |
| 208.7 | $I^-\cdot CH_2O_2\cdot 2H_2O$ | $CH_2O_2$ | 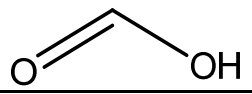 | Formic Acid |
| 216.5 | $I^-\cdot C_3H_6O_3$ | $C_3H_6O_3$ | 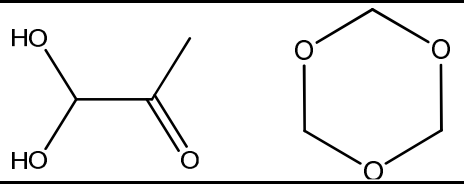 | Hydrated MG or cyclic F acetal |
| 230.3 | $I^-\cdot C_3H_4O_4$ | $C_3H_4O_4$ | 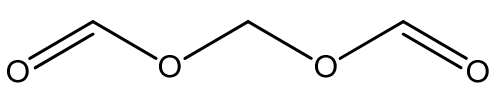 | n = 3 F hemiacetal |
| 252.4 | $I^-\cdot C_6H_6O_3$<br>$I^-\cdot C_3H_8O_4\cdot H_2O$<br>$I^-\cdot C_3H_6O_3\cdot 2H_2O$ | $C_6H_6O_3$<br>$C_3H_8O_4$<br>$C_3H_6O_3$ | 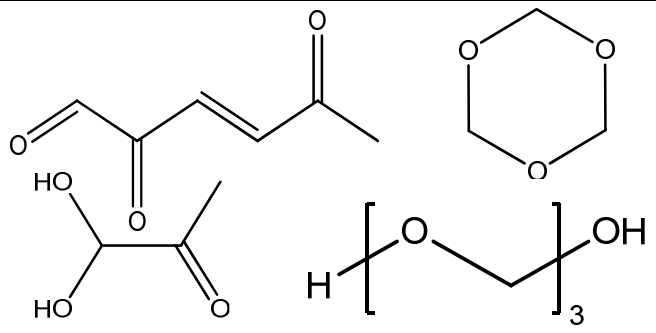 | MG aldol,<br>n = 3 F hemiacetal,<br>Hydrated MG, or cyclic F acetal |
| 257.4 | $C_8H_{17}O_9^-$ | $C_8H_{18}O_9$ | 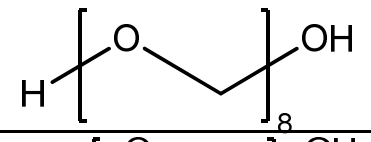 | n=8 F hemiacetal |
| 264.5 | $I^-\cdot C_4H_{10}O_5$ | $C_4H_{10}O_5$ | 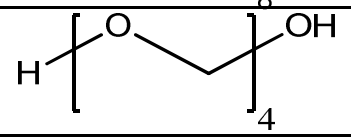 | n=4 F hemiacetal |
| 272.2 | $I^-\cdot C_6H_{10}O_4$ | $C_6H_{10}O_4$ | 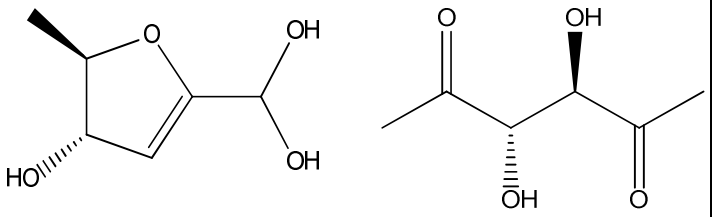 | MG aldol and hemiacetal |
| | $I^-\cdot C_5H_6O_5$ | $C_5H_6O_5$ | 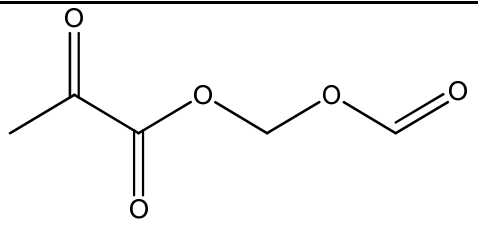 | MG + 2F hemiacetal |
| 288.1 | $I^-\cdot C_6H_{10}O_5$<br>$I^-\cdot C_6H_8O_4\cdot H_2O$<br>$I^-\cdot C_6H_6O_3\cdot 2H_2O$ | $C_6H_{10}O_5$<br>$C_6H_8O_4$<br>$C_6H_6O_3$ | 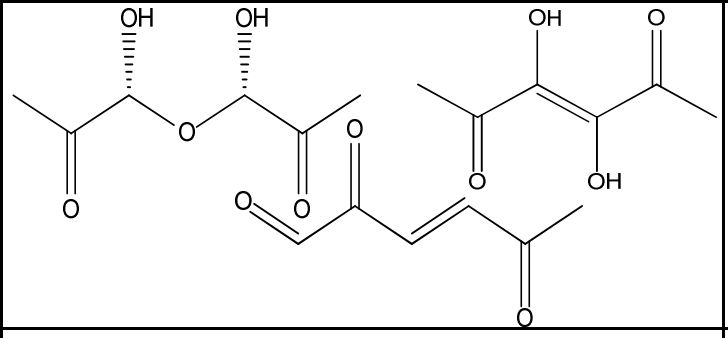 | MG aldol and hemiacetal |
| 314.3 | $I^-\cdot C_5H_{12}O_5\cdot 2H_2O$ | $C_5H_{12}O_5$ | 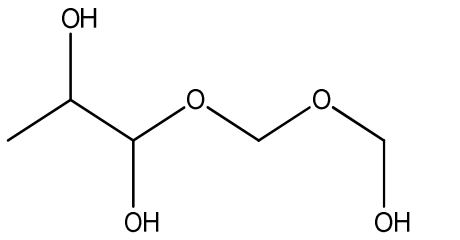 | MG + 2F hemiacetal |
| 324.5 | $I^-\cdot C_6H_{14}O_7$<br>$I^-\cdot C_6H_{12}O_6\cdot H_2O$<br>$I^-\cdot C_6H_{10}O_5\cdot 2H_2O$ | $C_6H_{14}O_7$<br>$C_6H_{12}O_6$<br>$C_6H_{10}O_5$ | 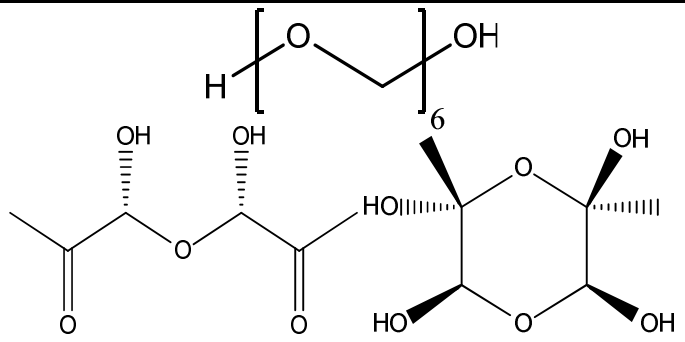 | n=6 F hemiacetal, MG hemiacetal |
| 342.6 | $I^-\cdot C_6H_{14}O_7\cdot H_2O$<br>$I^-\cdot C_6H_{12}O_6\cdot 2H_2O$ | $C_6H_{14}O_7$<br>$C_6H_{12}O_6$ | 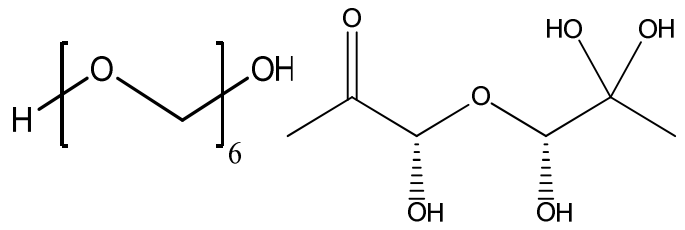 | n=6 F hemiacetal, MG hemiacetal |

**Table 4.** Proposed peak assignments for Aerosol-CIMS mass spectra with $H_3O^+$ of atomized solutions of 0.5 M acetaldehyde/MG (1:1) in 3.1 M AS

| m/z (amu) ± 1.0 amu | Ion Formula | Molecular Formula | Possible Structures | Mechanism |
|---|---|---|---|---|
| 84.4 | $CH_3O_2^+{\cdot}2H_2O$ | $CH_2O_2$ | 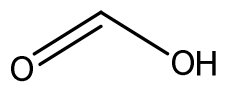 | Formic Acid |
| 88.9 | $C_3H_5O_3^+$ | $C_3H_4O_3$ | 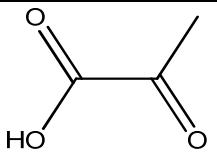 | Pyruvic Acid |
| | $C_4H_7O^+{\cdot}H_2O$<br>$C_4H_9O_2^+$ | $C_4H_6O$<br>$C_4H_8O_2$ | 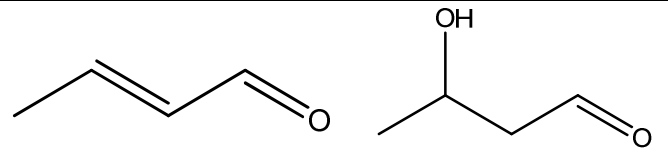 | A aldol |
| 93.5 | $C_2H_3O_3^+{\cdot}H_2O$ | $C_2H_2O_3$ | 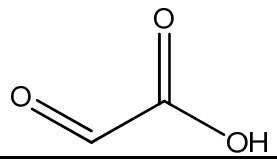 | Glyoxylic Acid |
| 95.5 | $C_2H_5O_3^+{\cdot}H_2O$ | $C_2H_4O_3$ | 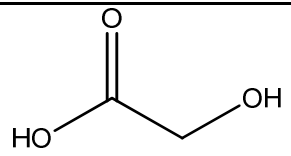 | Glycolic Acid |
| 98.4 | $C_2H_7O_2^+{\cdot}2H_2O$ | $C_2H_6O_2$ | 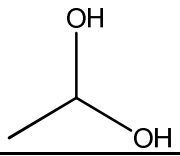 | Hydrated A |
| 107.2 | $C_3H_5O_3^+{\cdot}H_2O$ | $C_3H_4O_3$ | 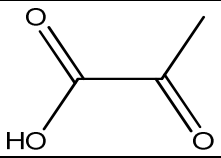 | Pyruvic Acid |
| | $C_4H_9O_2^+{\cdot}H_2O$ | $C_4H_8O_2$ | 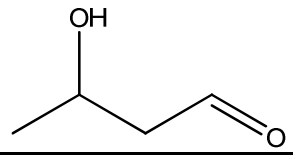 | A aldol |
| 126.0 | $C_7H_9O_2^+$ | $C_7H_8O_2$ | 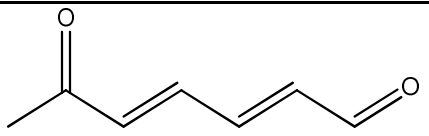 | MG + 2 A aldol |
| 134.0 | $C_5H_{11}O_4^+{\cdot}H_2O$ | $C_5H_{10}O_4$ | 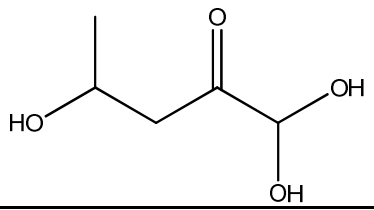 | MG + A aldol |
| 137.3 | $C_5H_{11}O_3^+{\cdot}H_2O$ | $C_5H_{10}O_3$ | 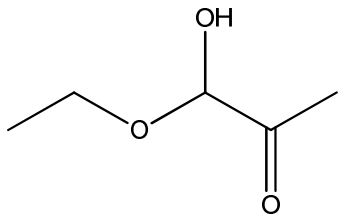 | Unknown |
| 145.1 | $C_6H_9O_4^+$<br>$C_6H_7O_3^+{\cdot}H_2O$ | $C_6H_8O_4$<br>$C_6H_6O_3$ | 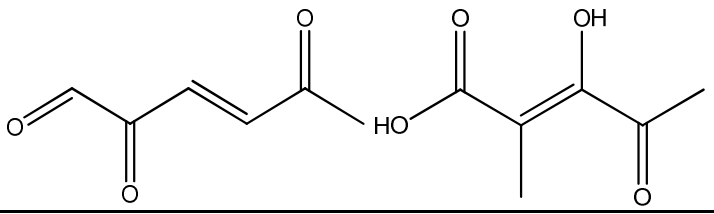 | MG aldol |
| 162.9 | $C_6H_{11}O_5^+$<br>$C_6H_9O_4^+{\cdot}H_2O$ | $C_6H_{10}O_5$<br>$C_6H_8O_4$ | 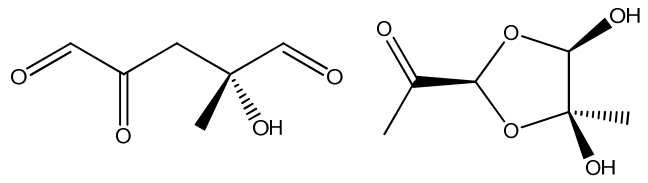 | MG hemiacetal and aldol |
| 164.7 | $C_6H_{13}O_5^+$ | $C_6H_{12}O_5$ | 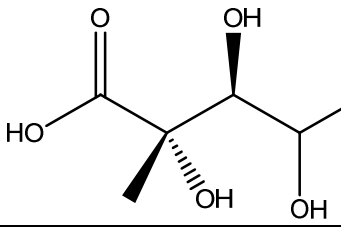 | MG aldol |
| | $C_6H_{11}O_4^+{\cdot}H_2O$ | $C_6H_{10}O_4$ | 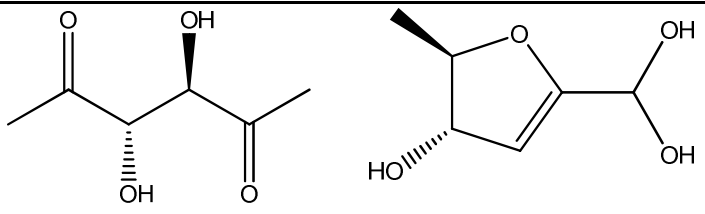 | MG hemiacetal and aldol |
| 192.9 | $C_{12}H_{15}O^+{\cdot}H_2O$ | $C_{12}H_{14}O$ | 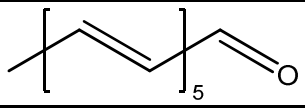 | 6 A aldol |
| 206.7 | $C_{11}H_{11}O_4^+$ | $C_{11}H_{10}O_4$ | 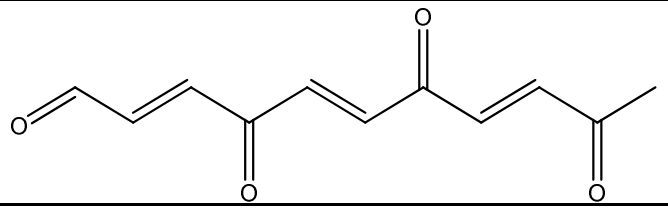 | A + 3 MG aldol |
| 235 | $C_9H_{15}O_7^+$ | $C_9H_{14}O_7$ | 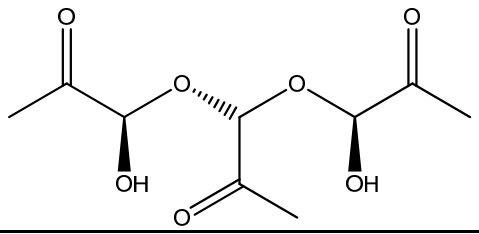 | MG hemiacetal |
| 248.9 | $C_{15}H_{17}O_2^+ \cdot H_2O$ | $C_{15}H_{16}O_2$ | 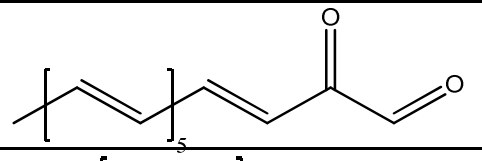 | 1MG + 6 A aldol |
| 289.6 | $C_{18}H_{21}O^+ \cdot 2H_2O$ | $C_{18}H_{20}O$ | 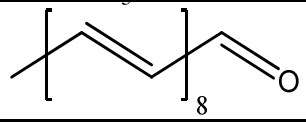 | 9 A aldol |
| 297 | $C_{20}H_{23}O^+ \cdot H_2O$ | $C_{20}H_{22}O$ | 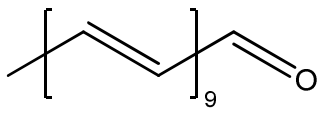 | 10 A aldol |

**Table 5.** Proposed peak assignments for Aerosol-CIMS mass spectra with $I^-$ of atomized solutions of 2 M acetaldehyde/MG (1:1) in 3.1 M AS.

| m/z (amu) ± 1.0 amu | Ion Formula | Molecular Formula | Possible Structures | Mechanism |
|---|---|---|---|---|
| 172.7 | $I^-{\cdot}CH_2O_2$<br>$I^-{\cdot}C_2H_6O$ | $CH_2O_2$<br>$C_2H_6O$ | 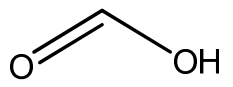 | Formic Acid |
| 186.4 | $I^-{\cdot}C_2H_4O_2$ | $C_2H_4O_2$ | 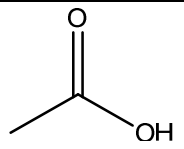 | Acetic Acid |
| 189.6 | $I^-{\cdot}C_2H_6O_2$ | $C_2H_6O_2$ | 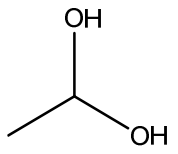 | Hydrated A |
| 194.6 | $C_6H_9O_6^-$<br>$C_2H_7O_6S^-{\cdot}H_2O$ | $C_6H_{10}O_6$<br>$C_2H_8O_6S$ | Unknown | Unknown |
| 208.4 | $I^-{\cdot}CH_2O_2{\cdot}2H_2O$ | $CH_2O_2$ | 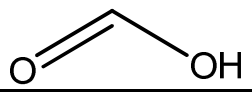 | Formic Acid |
| 216.3 | $I^-{\cdot}C_3H_6O_3$ | $C_3H_6O_3$ | 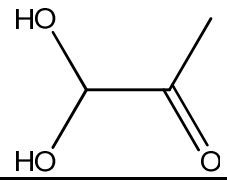 | Hydrated MG |
| 224.1 | $I^-{\cdot}C_2H_6O_2{\cdot}2H_2O$ | $C_2H_6O_2$ | 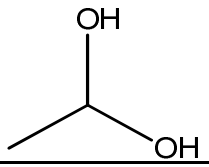 | Hydrated A |
| 230.7 | $I^-{\cdot}C_4H_8O_3$ | $C_4H_8O_3$ | 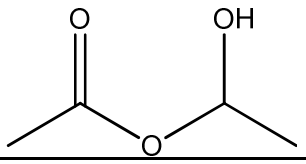 | A hemiacetal |
| | $I^-{\cdot}C_4H_6O_2{\cdot}H_2O$ | $C_4H_6O_2$ | 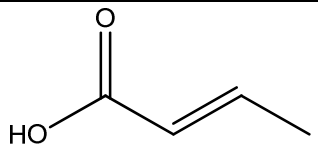 | Crotonic acid |
| 242.9 | $I^-{\cdot}C_5H_8O_3$ | $C_5H_8O_3$ | 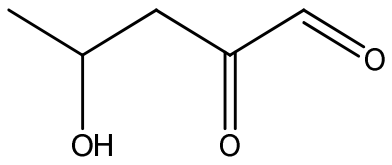 | MG + A aldol |
| 256.4 | $I^-{\cdot}C_6H_{10}O_3$ | $C_6H_{10}O_3$ | 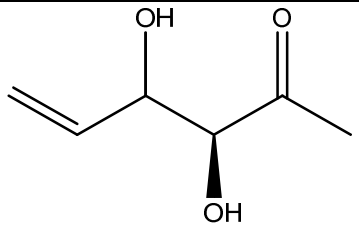 | MG aldol |
| 264.4 | $I^-{\cdot}C_4H_6O_3{\cdot}2H_2O$ | $C_4H_6O_3$ | 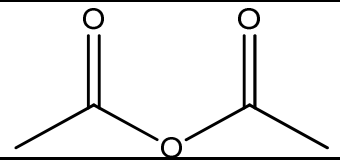 | A hemiacetal |
| 269.5 | $I^-{\cdot}C_4H_{10}O_3{\cdot}2H_2O$ | $C_4H_{10}O_3$ | 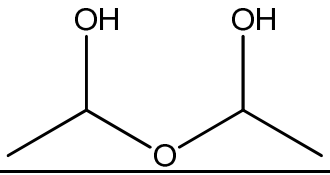 | A hemiacetal |
| 272.2 | $I^-{\cdot}C_6H_{10}O_4$ | $C_6H_{10}O_4$ | 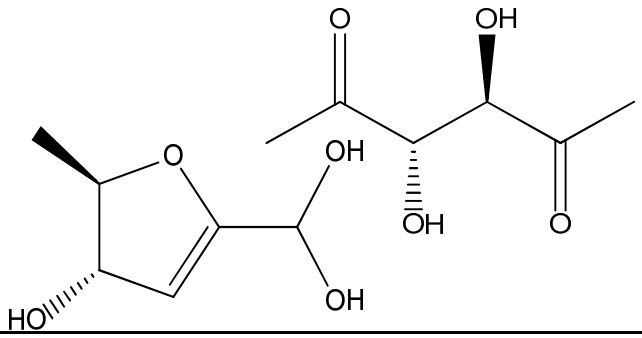 | MG aldol and hemiacetal |
| 342.3 | $I^-{\cdot}C_6H_{12}O_6{\cdot}2H_2O$ | $C_6H_{12}O_6$ | 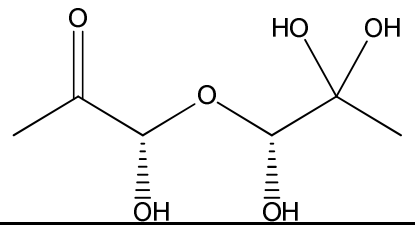 | MG hemiacetal |

## *Figures*

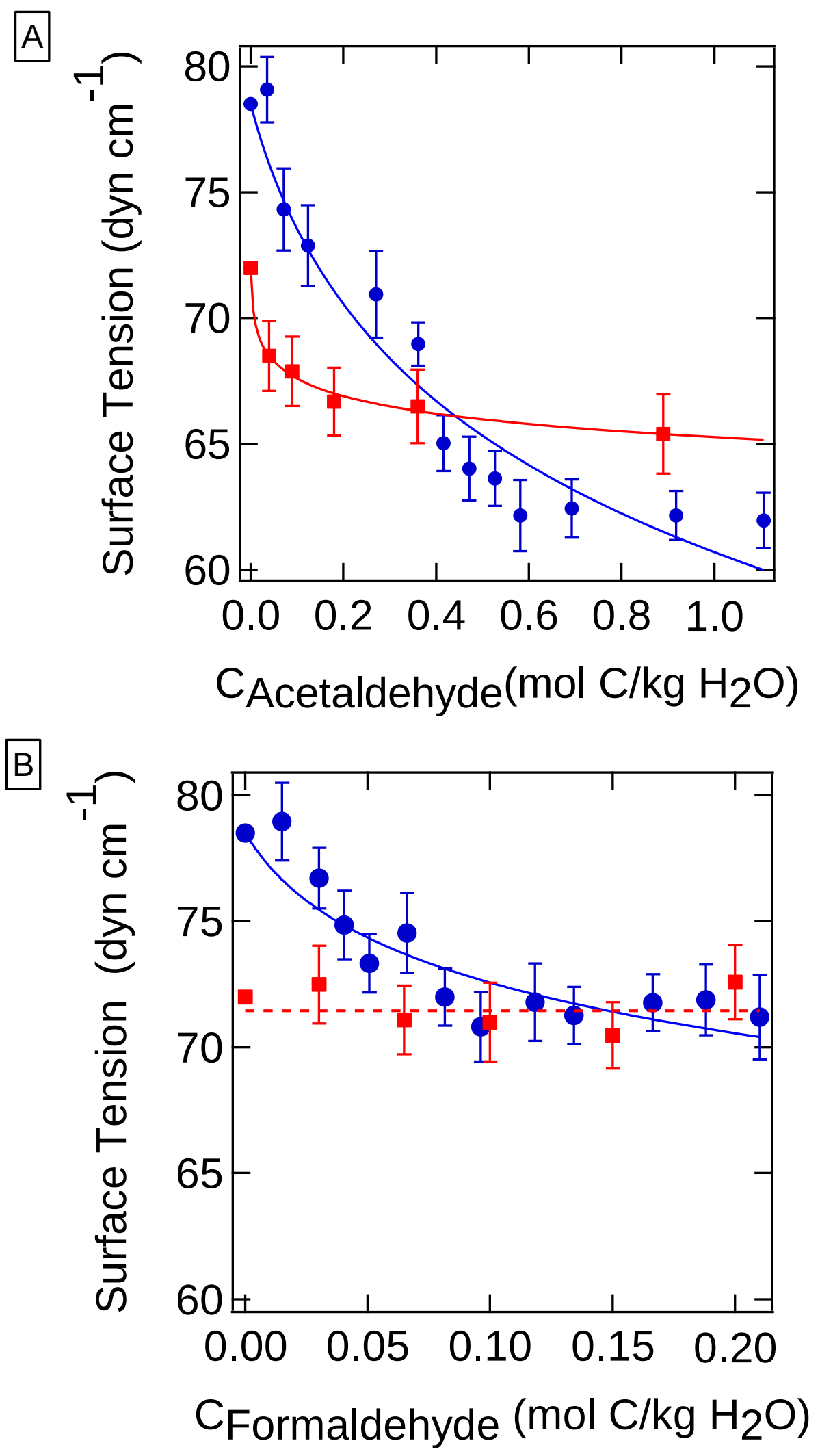

**Figure 1.** Surface tension of solutions containing (A) acetaldehyde and (B) formaldehyde in 3.1 M AS (●) and in water (■). The curves shown are fits to the data using the Szyszkowski-Langmuir equation (Eq. (2)). A linear fit (red dashed line) is shown for the formaldehyde-water data as a guide to the eye.

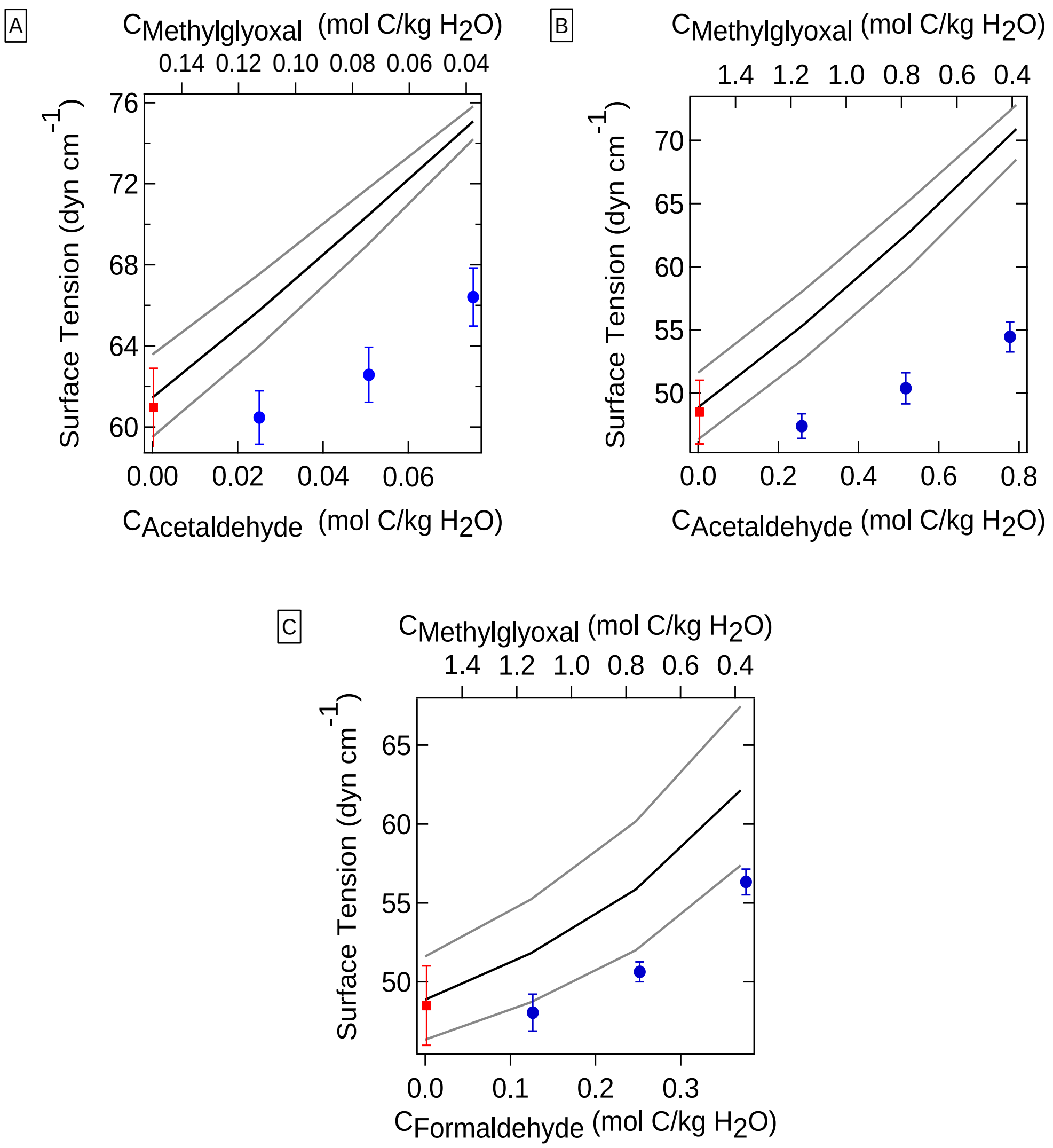

**Figure 2**. Surface tension of binary mixtures of acetaldehyde or formaldehyde with MG in 3.1 M AS solutions. The total organic concentration was 0.05 M (A) or 0.5 M (B, C). The black line shows Henning model predictions (Eq. (3)) using the parameters listed in Table 1. The grey lines show the confidence interval of the model predictions. ■: MG in AS (based on the Szyszkowski-Langmuir equation (Eq. (2)), using the parameters in Table 1). ●: Acetaldehyde (A and B) or Formaldehyde (C) with MG in 3.1 M AS solutions.

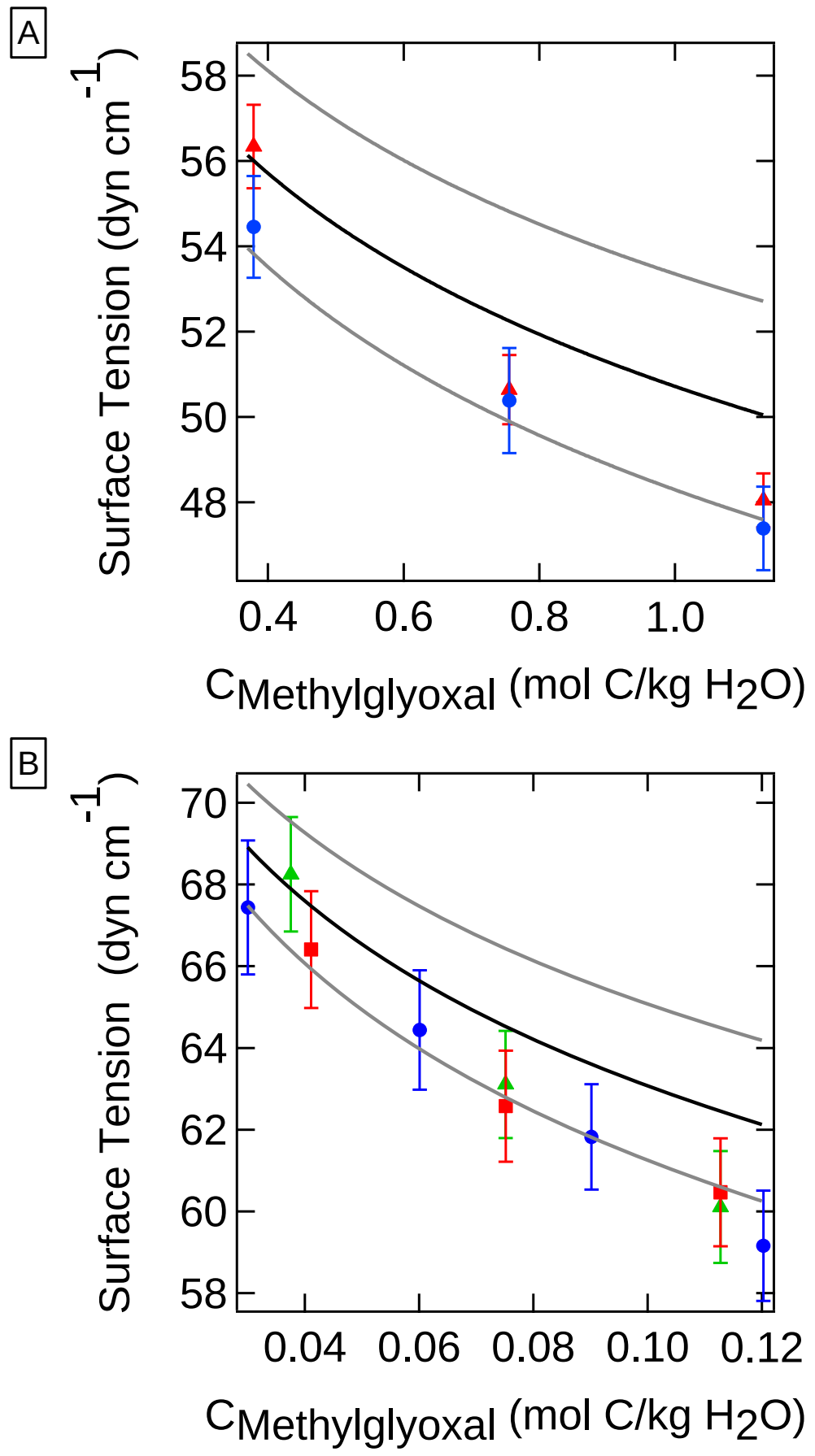

**Figure 3**. Surface tension in binary and ternary organic mixtures (Fig 2 & 3) as a function of MG concentration. A) Binary mixtures (0.5 M total organic concentration) ▲: acetaldehyde-MG, ●: formaldehyde-MG B) 0.05 M total organic concentration. ▲: ternary mixture (acetaldehyde:formaldehyde=1:1 by mole, varying MG); ●: ternary mixture (acetaldehyde:formaldehyde=1:3 by mole, varying MG); ■: binary mixture (acetaldehyde-MG). Black curves indicate the Szyszkowski-Langmuir curve for MG in AS using the parameters in Table 1. Grey curves show the confidence intervals.

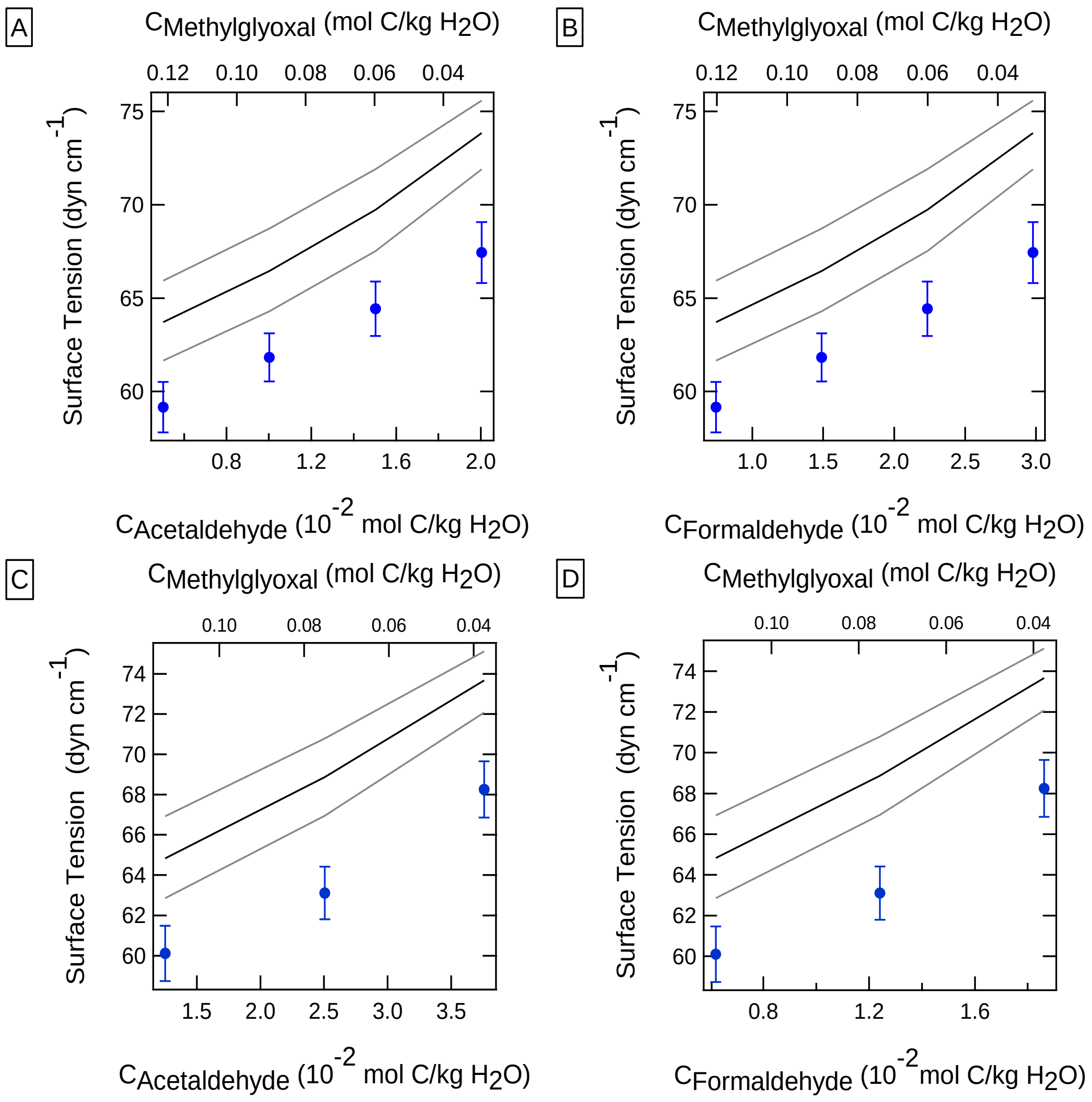

**Figure 4**. Surface tension data for ternary (acetaldehyde, formaldehyde and MG) mixtures in 3.1 M AS solutions. The molar ratios of acetaldehyde to formaldehyde are 1:3 (A and B) and 1:1 (C and D). The total organic concentration was constant at 0.05 M. The black line shows Henning model predictions using the parameters listed in Table 1. The grey lines show the confidence interval of the predicted data.

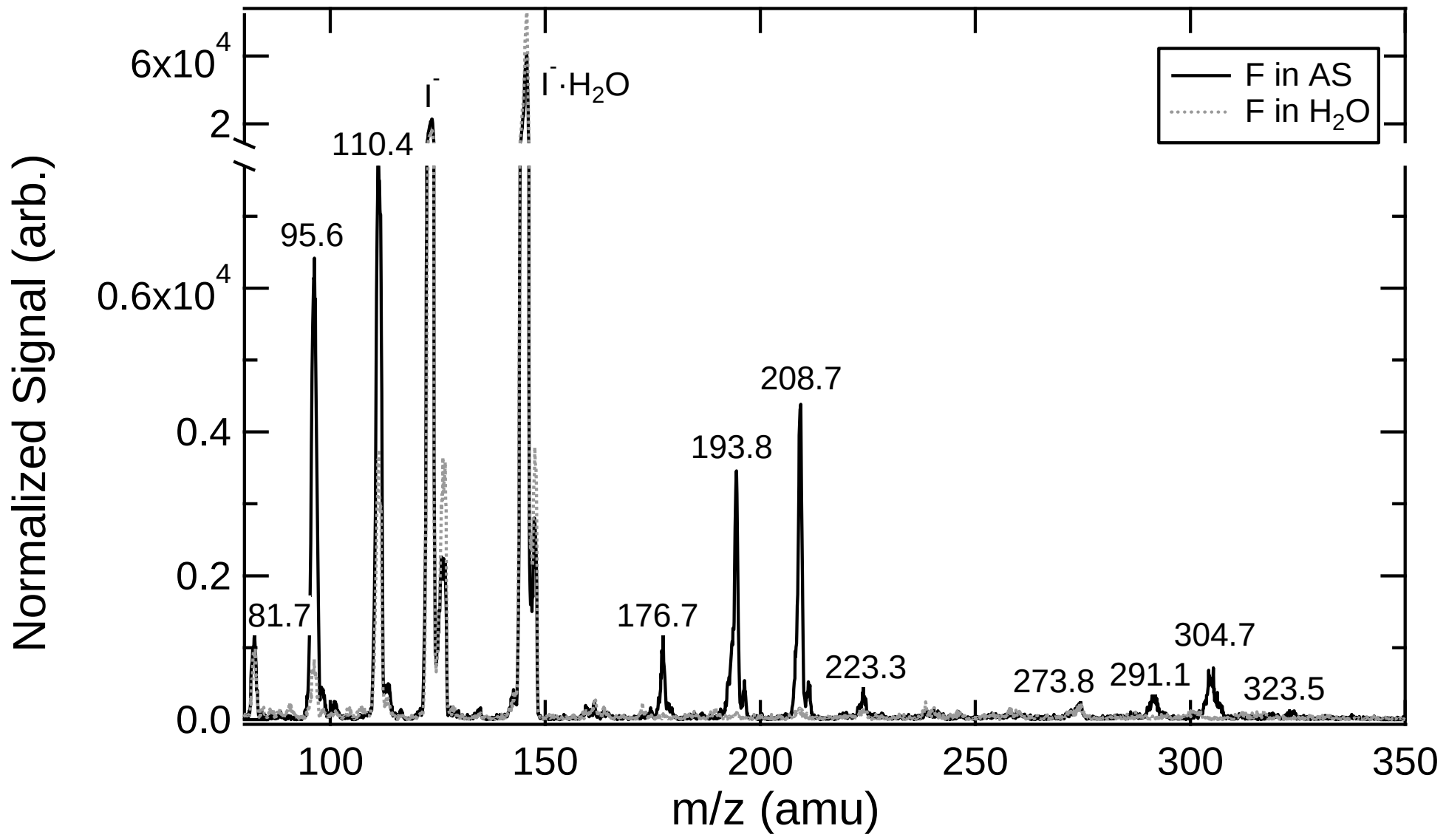

**Figure 5.** Aerosol-CIMS spectra of atomized solutions of 0.2 M formaldehyde in 3.1 M AS and $H_2O$. See the text for details of sample preparation and analysis. Negative-ion mass spectrum obtained using $I^-$ as the reagent ion.

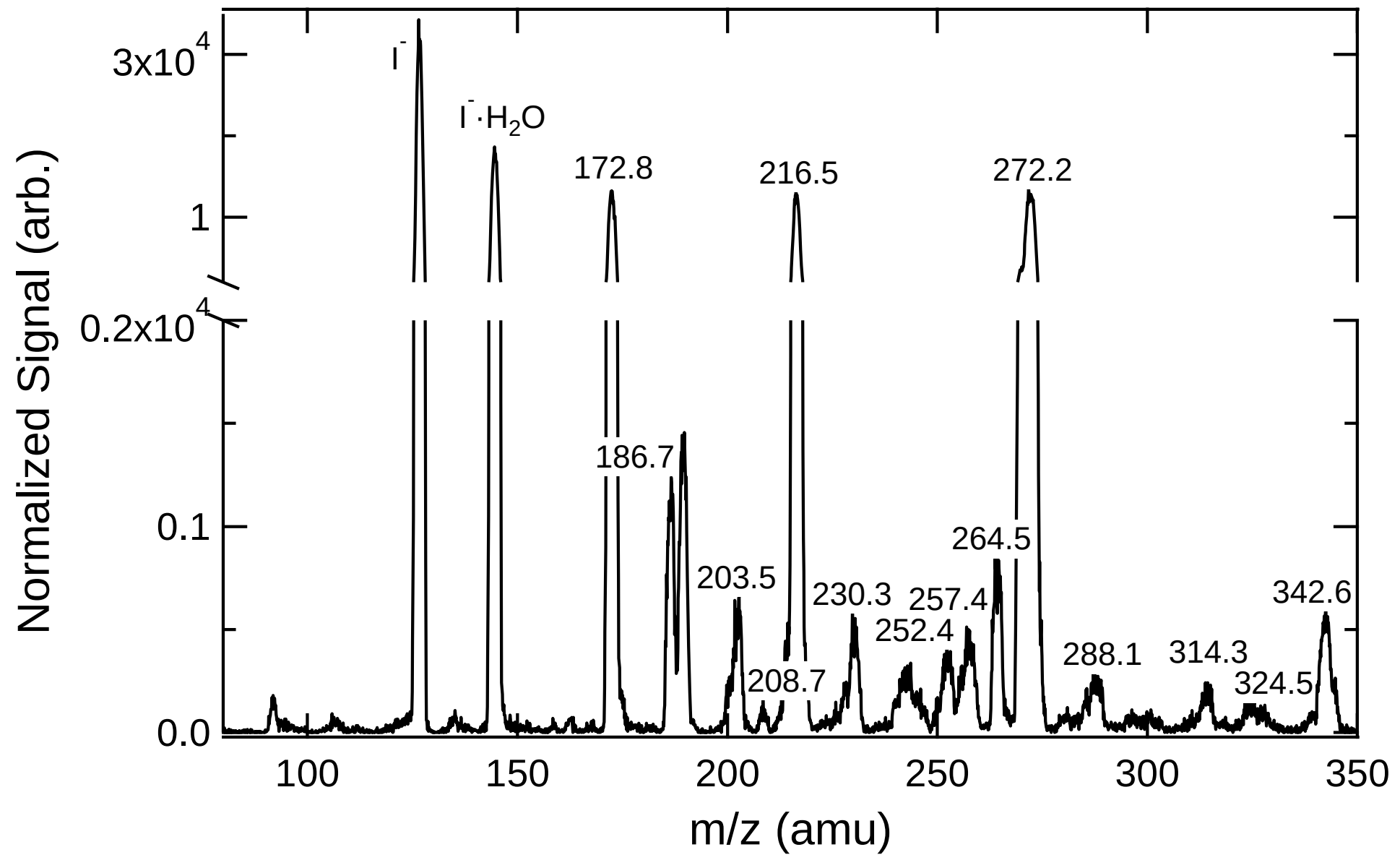

**Figure 6.** Aerosol-CIMS spectra of atomized solutions of 2 M formaldehyde/MG (1:1) in 3.1 M AS. See the text for details of sample preparation and analysis. Negative-ion mass spectrum obtained using $I^-$ as the reagent ion.

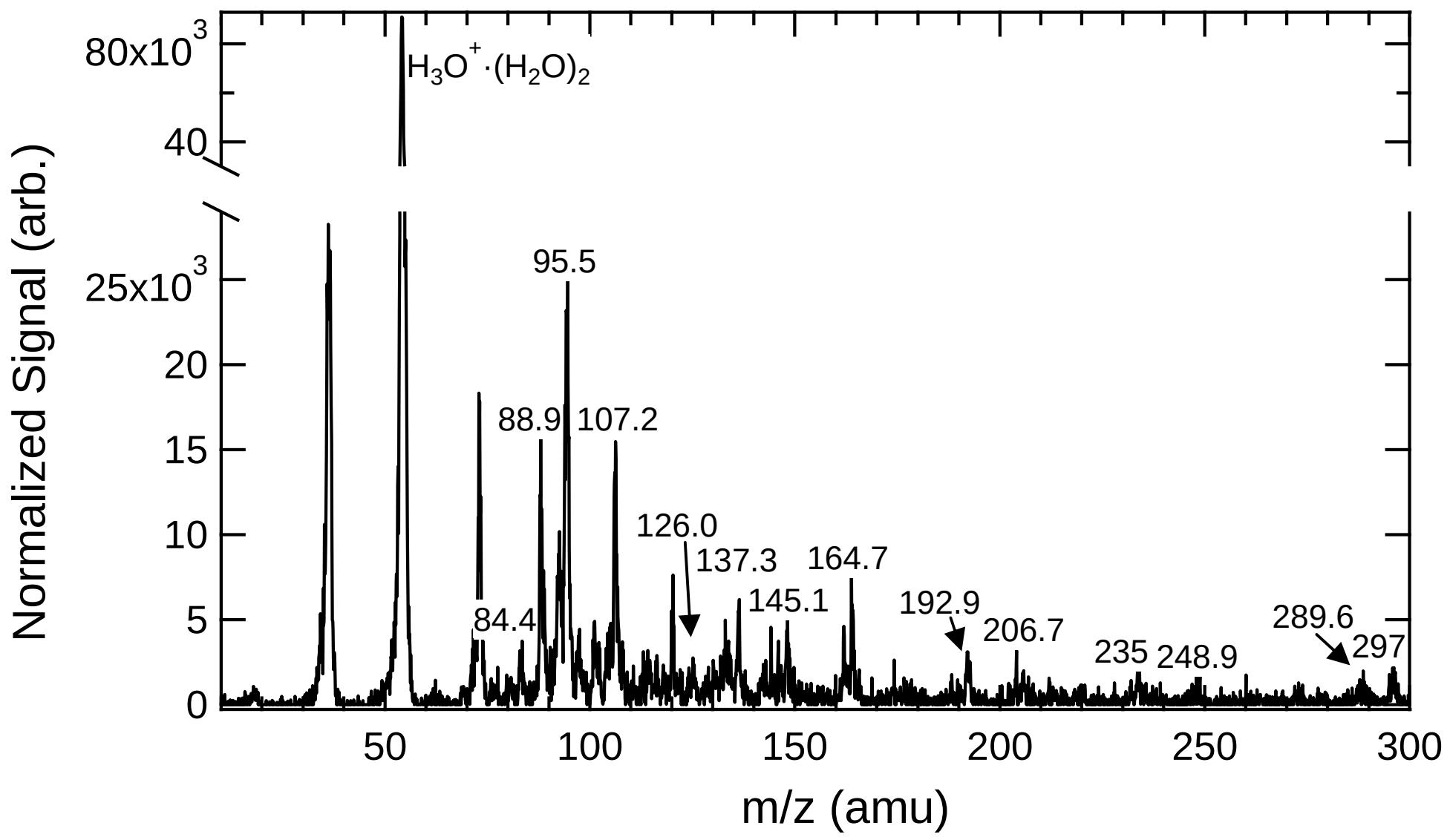

**Figure 7.** Aerosol-CIMS spectra of atomized solutions of 0.5 M acetaldehyde/MG (1:1) in 3.1 M AS. See the text for details of sample preparation and analysis. Positive-ion mass spectrum using $H_3O^+{\cdot}(H_2O)_n$ as the reagent ion.

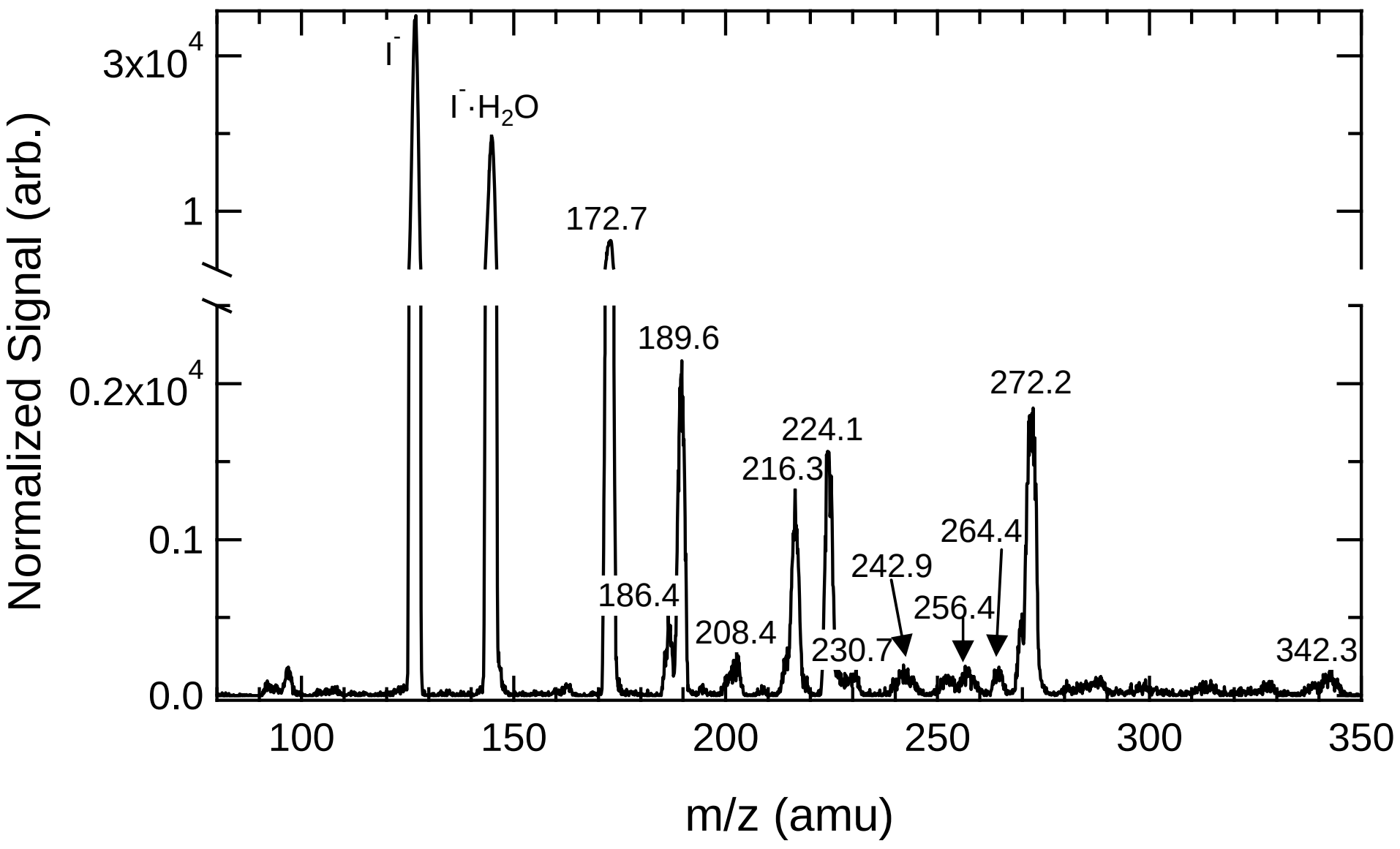

**Figure 8.** Aerosol-CIMS spectra of atomized solutions of 2 M acetaldehyde/MG (1:1) in 3.1 M AS. See the text for details of sample preparation and analysis. Negative-ion mass spectrum obtained using $I^-$ as the reagent ion.